\documentclass[twocolumn,superscriptaddress,showpacs,prl]{revtex4-1}    
\usepackage{graphicx,amsmath,mathrsfs}
\usepackage{amssymb}
\usepackage{amsthm}
\usepackage{bm}
\usepackage{subfigure,xcolor}   

\def\bc{\begin{center}}
\def\ec{\end{center}}

\newcommand{\bs}[1]{\boldsymbol{#1}}

\newcommand{\pd}{{\phantom{\dag}}}
\newcommand{\up}{\uparrow}
\newcommand{\dw}{\downarrow}
\newcommand{\eps}{\varepsilon}

\def\ie{\emph{i.e.},\ }
\def\eg{\emph{e.g.}\ }
\def\ea{\emph{et al.\,\,}}

\begin{document}
\title{Quantum paramagnet in a $\bs{\pi}$ flux triangular lattice
  Hubbard model}
\author{Stephan Rachel}
\affiliation{Institute for Theoretical Physics, Technische Universit\"at Dresden, 01062 Dresden, Germany}
\author{Manuel Laubach}
\affiliation{Institute for Theoretical Physics, University of W\"urzburg, 97074 
W\"urzburg, Germany}
\author{Johannes Reuther}
\affiliation{Dahlem Center for Complex Quantum Systems and Fachbereich
  Physik, Freie Universit\"at Berlin, 14195 Berlin, Germany}
\affiliation{Helmholtz-Zentrum Berlin f\"ur Materialien und Energie, 14109 Berlin, Germany}
\author{Ronny Thomale}
\affiliation{Institute for Theoretical Physics, University of W\"urzburg, 97074 
W\"urzburg, Germany}

\date{\today}

\begin{abstract}
  We propose the $\pi$ flux triangular lattice Hubbard model
  ($\pi$-THM) as a prototypical setup to stabilize magnetically
  disordered quantum states of matter in the presence of charge
  fluctuations. 
The quantum paramagnetic domain of the $\pi$-THM which we identify for
intermediate Hubbard~$U$ is framed by a Dirac semi-metal for weak
coupling and by 120${}^{\circ}$ N\'eel order for strong coupling.
Generalizing the
  Klein duality from spin Hamiltonians to tight-binding models, the
  $\pi$-THM maps to a Hubbard model which corresponds to the
  $(J_{\text{H}},J_{\text{K}})=(-1,2)$ Heisenberg-Kitaev model in its strong coupling
  limit. The $\pi$-THM provides a promising microscopic testing ground for
  exotic finite-$U$ spin liquid ground states amenable to numerical
  investigation.
\end{abstract}


\maketitle

{\it Introduction.} 
Two-dimensional quantum paramagnets such as spin liquids or valence
bond crystals are quantum states of matter that, albeit their
enormous diversity from a theoretical
standpoint, are hard to find in experimental
scenarios~\cite{diep04,lacroix,lees08,balentsn10}. At the
level of theoretical identification in microscopic models, recent numerical advances
such as two-dimensional density matrix renormalization
group~\cite{whites11,jiang-j1j2,depenbrock-12prl067201},
pseudofermion renormalization group~\cite{suttner-kagome}, or variational Monte Carlo~\cite{PhysRevB.72.045105,clark-11-j1j2}
could provide substantiated support for spin liquid
regimes. 
Predominantly, the strong coupling limit of a Mott state is
considered which is parametrized by spin exchange interactions.

An 
exception constitutes the work by Meng \ea on the Hubbard
model for the honeycomb lattice, where quantum Monte Carlo algorithms avoid the sign
problem due to lattice bipartiteness~\cite{meng-10n847}.
Small system sizes suggested a non-magnetic insulating regime
without valence bond crystal order~\cite{meng-10n847}. As larger scale calculations~\cite{sorella-12sr992}
and more refined determinations of the order parameter~\cite{PhysRevX.3.031010,schmidt2014} revealed,
however, the metal insulator transition turns out to be of Gross-Neveu type, where antiferromagnetic order sets in
immediately. This is confirmed by cluster methods operating at intermediate Hubbard~$U$~\cite{PhysRevLett.110.096402,PhysRevB.90.165136}. What
can still be taken as a motivation from this finding is that a Dirac metal
for weak coupling might contribute to a promising scenario for an unconventional metal-insulator transition and exotic phases at intermediate coupling, which is the starting point of
our analysis. 

%

In this Letter, we propose the Hubbard model on the $\pi$ flux triangular lattice
($\pi$-THM) as a prototypical candidate for quantum paramagnetic phases at
intermediate coupling. In its weak coupling
  limit, the band structure of the $\pi$-THM is semi-metallic,
  exhibiting the same low-energy behavior as graphene with a different
  Fermi velocity. The stability of this semi-metal with respect to
  weak local Coulomb interactions follows from generic properties of
  Dirac electrons~\cite{abrikosov,herbut06prl146401}. In its
  strong-coupling limit, positive and negative hoppings give rise to
  the same spin exchange amplitude $J=4(\pm t)^2/U$, rendering the
  $\pi$-THM identical to the nearest-neighbor Heisenberg model
  on the triangular lattice yielding 120$^{\circ}$
N\'eel order~\cite{PhysRevLett.82.3899}. Generalizing the Klein duality~\cite{khaliullin05ptps155,jackeli-09prl017205,reuther-12prb155127,kimchi-14prb014414} from spin models
to tight-binding models, we can relate the $\pi$-THM to a transformed
Hubbard model with bond-selective Kitaev-like hopping amplitudes. The strong coupling limit of this Klein-transformed model is given by the $(J_{\text{H}},J_{\text{K}})=(-1,2)$ Heisenberg-Kitaev model.

As the weak and strong
coupling limits are fixed, it remains to be investigated whether there
is a direct semi-metal to magnet transition, or whether an
intermediate paramagnetic domain emerges at the metal insulator transition. 
We will explicate below that perturbing away from its infinite-$U$ limit, the short-range resonating valence bond
(RVB) loops lower the energy in the $\pi$-THM more significantly than
for the regular Hubbard model on the triangular lattice (THM). These perturbative arguments are supplemented by
the calculation of single-particle spectral functions and ordering tendencies via
variational cluster approximation (VCA). For intermediate Hubbard $U$,
we find an extended non-magnetic insulating regime, which promises to
host unconventional quantum paramagnetic states of matter.

\begin{figure}[t]
\centering
\includegraphics[scale=0.47]{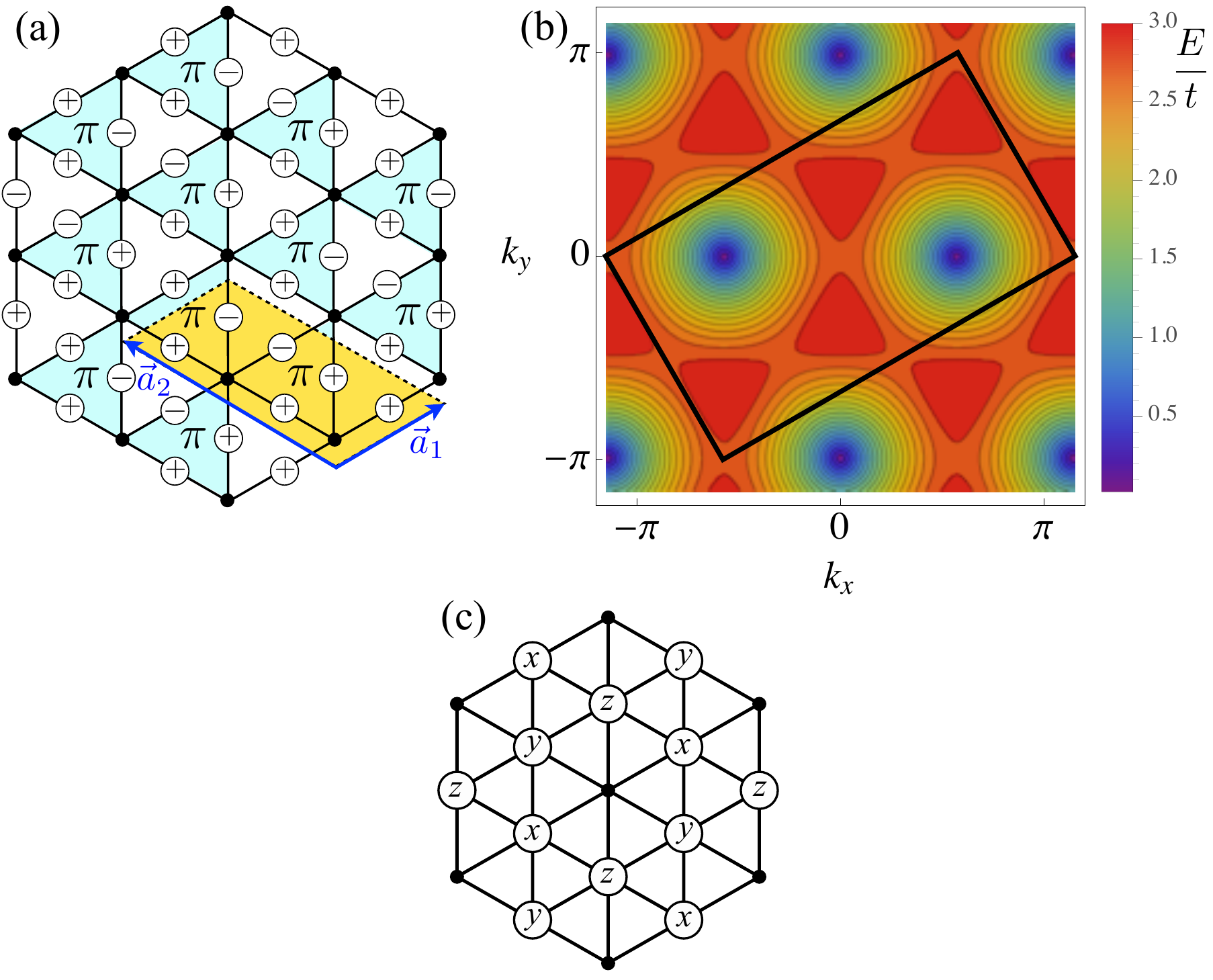}
\caption{(Color online) a) Flux pattern and signs of the real nearest
  neighbor hoppings on the triangular lattice. The $\pi$'s denote
  triangles which are  threaded by a $\pi$ flux. b) Contour plot of the spin-degenerate
  upper band of the semi-metallic band structure (blue spots
  indicate the Dirac nodes at zero energy); the lower band follows
  from particle-hole symmetry. The Brillouin zone (black) is spanned
  by the reciprocal vectors
  $\bs{b}_1=\frac{2\pi}{\sqrt{3}}(1,\sqrt{3})$ and
  $\bs{b}_2=\frac{2\pi}{\sqrt{3}}(-1/2, \sqrt{3}/2)$. (c)
  four-sublattice structure of a Klein transformation. 
  ``x", ``y" and ``z" indicate the axis around which a spin rotation
  of angle $\pi$ is performed. The full dot sublattice remains unchanged.}  
\label{fig:lattice}
\end{figure}

%
%
{\it $\pi$-THM.} 
We consider a triangular lattice with an alternating flux pattern such
that a triangle threaded by a $\pi$ flux is surrounded by three
triangles which are flux-free (Fig.~\ref{fig:lattice}a). Note that triggered by the
success in formulating and manufacturing flux lattices through artificial gauge fields in
ultra-cold atomic
gases~\cite{jaksch-03njp56,dalibard-11rmp1523,cooper}, flux patterns in lattice models have become an
experimentally relevant topic in contemporary AMO physics.
Of particular importance is the recent proposal of  ``shaking'' an
optical lattice, \ie to apply a periodic force pulse to the optical
lattice. It has been shown that staggered flux patterns for triangular
lattices can be achieved\,\cite{struck-12prl225304}. As the Hubbard
coupling can be naturally adjusted by the optical setup, this might be
one promising route towards realizing the model subject to this paper.

In Fig.~\ref{fig:lattice}a, the signs of the real nearest-neighbor
hoppings $t_{ij}$ are shown which reproduce the alternating flux pattern; the  two-atomic unit cell is shown in yellow spanned by the primitive vectors $\bs{a}_1=(\sqrt{3}/2,1/2)$ and $\bs{a}_2=(-\sqrt{3},1)$. (We set the lattice spacing $a\equiv 1$ throughout the paper.)
 The band structure is spin-degenerate and particle-hole symmetric. The Bloch matrix reads $h(\bs{k})=\sum_{\bs{k}} \bs{d}(\bs{k})\cdot \bs{\sigma}$,
where $\sigma^\alpha$ are Pauli matrices related to the sublattice degree of freedom and
\begin{equation}
\bs{d}(\bs{k})=t \left(\begin{array}{c} 1+\cos{\bs{a}_1 \bs{k}}+\cos{\bs{a}_2 \bs{k}}-\cos{(\bs{a}_1+\bs{a}_2) \bs{k}}\\[5pt]
\sin{\bs{a}_1 \bs{k}}-\sin{\bs{a}_2 \bs{k}}+\sin{(\bs{a}_1+\bs{a}_2) \bs{k}}\\[5pt]
 2 \cos{\bs{a}_1\bs{k}}
\end{array}\right).\ 
\end{equation}
We find the single particle spectrum $\eps^\pm_\sigma = \pm  \sqrt{
  d^2 }$ which is shown as a contour plot in
Fig.\,\ref{fig:lattice}b. The Dirac nodes are located at $K^\pm =
(\pm\pi/\sqrt{3},0)$. Expanding around the Dirac nodes yields the
Lorentz-invariant Dirac theory of graphene, along with valley and spin
degeneracy and a bare Fermi-velocity $v_F = \sqrt{6} t$.
Note that this type of band structure has been
previously mentioned as a mean field spectrum of a U(1) liquid candidate (dubbed U1B)\,\cite{wen-unpub}.
Adding Hubbard interactions, we find the $\pi$-THM governed by the Hamiltonian
\begin{equation}\label{ham:piTHM}
\mathcal{H}_{\pi\text{-THM}}=\sum_{\langle ij \rangle,\sigma} \left( t_{ij} c_{i\sigma}^\dag c_{j\sigma}^\pd + {\rm h.c.}\right) + U\sum_i n_{i\up} n_{i\dw},
\end{equation}
where $n_{i\sigma}=c_{i\sigma}^\dag c_{i\sigma}^\pd$ denotes the
density operator of electrons at site $i$ with spin $\sigma$.

%
%
{\it Klein-duality map from a Kitaev-Hubbard model.} 
The $\pi$ flux pattern on the triangular lattice allows to draw subtle
connections to iridium-based transition metal oxides and
Heisenberg-Kitaev models\,\cite{jackeli-09prl017205}. Originally
proposed for the honeycomb iridates
(Na,Li)$_2$IrO$_3$\,\cite{jackeli-09prl017205}, the Heisenberg-Kitaev model reads
\begin{equation}\label{hk-model}
\mathcal{H}_{\rm HK}=\sum_{\langle ij \rangle} J_{\text{H}} \bs{S}_i\bs{S}_j + J_{\text{K}} S_i^\gamma S_j^\gamma,
\end{equation}
where, for the triangular lattice, we define $\gamma=x$ for bonds along the $\bs{a}_1$ direction, $\gamma=y$ along the $\bs{a}_2$ direction, and $\gamma=z$ along the vertical bonds.

We define the class of Kitaev-Hubbard models as tight-binding band structures
subject to local Hubbard $U$
which, in the limit of infinite $U$, map onto a
Heisenberg-Kitaev model~\eqref{hk-model}.  As explicated below, we can formulate a Klein duality
map from our $\pi$-THM to such a Kitaev-Hubbard model.
In the past, Klein dualities have been successfully applied to spin
Hamiltonians\,\cite{khaliullin05ptps155,jackeli-09prl017205,reuther-12prb155127,kimchi-14prb014414}. Here,
we generalize the Klein duality to Hubbard models, \ie to creation and
annihilation operators of electrons. As for the spin models, we define
four sublattices on the triangular lattice as shown in Fig.\,\ref{fig:lattice}c,
and then rotate the spin of the creation/annihilation operators on the
different sublattices: we rotate the first sublattice around the
$x$ axis by $\pi$, the second around the $y$ axis by
$\pi$, the third around the $z$ axis by $\pi$, and the fourth sublattice remains unchanged (Fig.~\ref{fig:lattice}c).
Such spin rotations are easily accomplished by virtue of Pauli matrices, $\mathcal{U}_\alpha = \exp{(i\frac{\pi}{2} \sigma^\alpha)} 
= i \sigma^\alpha$ and ${\mathcal{U}^\alpha}^\dag = -i \sigma^\alpha$
for rotations around the $\alpha$ axis ($\alpha=x,y,z$). The Klein-transformed version of Eq.~(\ref{ham:piTHM}) is given by a Kitaev-Hubbard model with a kinetic term $H_0 =
it\sum_{\langle ij \rangle} \nu_{ij} c_{i\alpha}^\dag
\sigma^\gamma_{\alpha\beta} c_{j\beta}^\pd$ where $\gamma=x,y,z$ are
defined as in \eqref{hk-model}, and $\sigma^\gamma$ denote Pauli
matrices describing the spin degrees of freedom. The phase convention
$\nu_{ij}=\pm 1$ is chosen such that the hopping amplitude is positive
in $\bs{a}_1$, $\bs{a}_2$ and $(2\bs{a}_1+\bs{a}_2)$ directions and
negative in opposite directions. The Hubbard term $U n_{i\up}
n_{i\dw}$ is invariant under the Klein map. In the strong
coupling limit, this Kitaev-Hubbard model becomes the $(J_{\text{H}},J_{\text{K}})=(-1,2)$
Kitaev-Heisenberg model of Eq.~(\ref{hk-model}).

Depending on the specific values of $J_{\text{H}}$ and $J_{\text{K}}$,
the model in Eq.~(\ref{hk-model}) might give rise to nonmagnetic phases of triangular
layered iridate compounds\,\cite{PhysRevB.86.140405,trebst2017,li-14arxiv1409.7820}. From
the classical analysis, the $(J_{\text{H}},J_{\text{K}})=(-1,2)$ point
in Eq.~(\ref{hk-model}) is surrounded by a $\mathbb{Z}_2$ vortex
lattice\,\cite{rousochatzakis-12arXiv:1209.5895}. As such, the Klein
map suggests that charge fluctuations in the
$\pi$-THM, as implied by finite $U$, may trigger similarly interesting effects as for the vortex lattice.

{\it Variational Cluster Approximation.} VCA is a quantum cluster
approach to compute single-particle spectral functions for interacting
many-body systems~\cite{potthoff03epjb335}. One first
solves a small cluster (with typically four to twelve lattice sites)
exactly and derives the corresponding full Green's function. Using the
framework of self-energy functional theory\,\cite{potthoff03epjb429},
one obtains the full Green's function of an infinitely large lattice
which is covered by the clusters, and the individual clusters
are coupled by hopping terms only. The latter step represents a
significant approximation to the full many-body problem, while the
method still includes spatial quantum correlations. Embedded into a
grand-canonical ensemble, the configuration with lowest energy
is found by varying the chemical potential as well as all
single-particle parameters which may also include the bare hopping
amplitudes~\cite{PhysRevB.90.165136}. 
Three comments about the VCA are in order which are important for the
analysis of our $\pi$-THM. (i) Variation of the hopping $t$ is crucial
in order to guarantee the stability of the semi-metal with respect to
small $U$. 
(ii) Using the same setup and accuracy which due to (i) exceeds
most previous VCA calculations, we do not find a nonmagnetic
insulator (NMI) phase within the honeycomb lattice Hubbard model for
intermediate $U$~\cite{PhysRevB.90.165136} as a benchmark.
(iii) The magnetic instability is investigated by
means of Weiss fields. For the 120$^\circ$ N{\'e}el order, only clusters
with multiples of three lattice sites can be used. In conjunction with
the two-atomic unit cell, only 6 and 12 site clusters are in principle
suitable for the analysis of the $\pi$-THM. Preference is of course
given to 12, \ie the
largest available cluster.

\begin{figure}[t]
\!\!\!\includegraphics[width=1.07\linewidth]{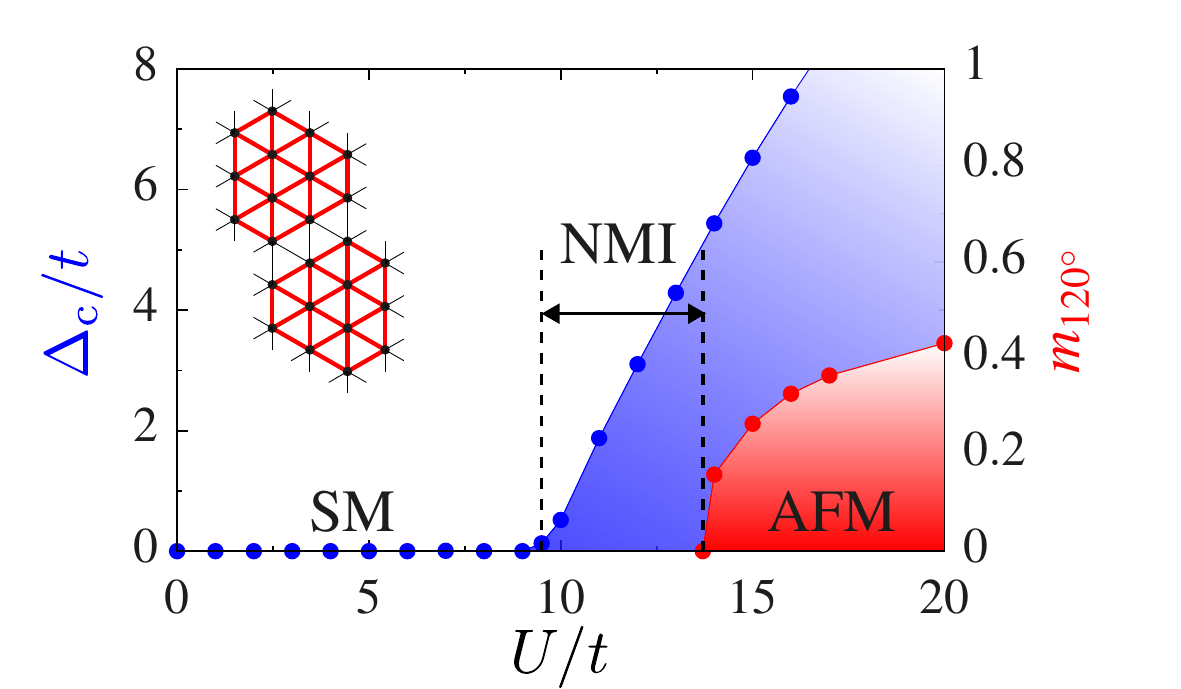}
\caption{(Color online). Phase diagram of~\eqref{ham:piTHM} as obtained by VCA. $U_{c1}$, $U_{c2}$, $\Delta_{\rm c}$, and
  $m_{120^\circ}$ are calculated for a lattice covering with 12-site
  (mirror) clusters as sketched in the inset (the magnetization $m_{120^\circ}=1$ denotes the saturation value). From single-particle
  spectra, there are three
  phases: semi-metal (SM), non-magnetic insulator (NMI), and 120${}^{\circ}$ N\'eel antiferromagnetic
  insulator (AFM). }
\label{fig:spec2vca}
\end{figure}

\begin{figure*}[t]
\includegraphics[width=0.99\linewidth]{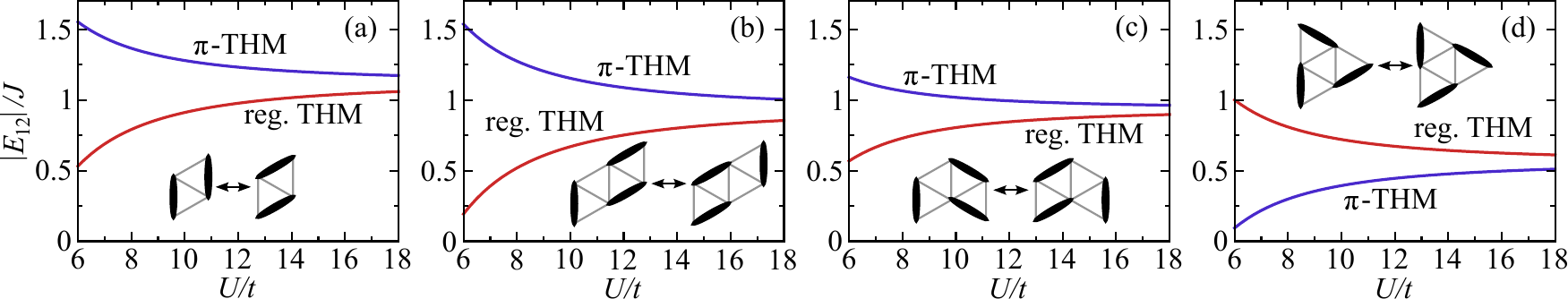}
\caption{(Color online). Strength of dimer loop resonances $|E_{12}|$
  (see Eqs. \ref{resonance} and~\ref{singlet}) for the 4-site loop in
  panel (a) and all 6-site loops in panels (b) - (d) calculated for
  the $\pi$-THM and the regular THM. The insets illustrate the
  according resonances. Except for (d), dimer resonances result in a
  larger energy gain for the $\pi$-THM than the regular THM.}
\label{fig:dimer}
\end{figure*}

%
%
{\it Phase diagram.} 
For the quantitative analysis, we employ 
a super-cluster construction with a 12-site and
mirror-12-site cluster (see inset Fig.~\ref{fig:spec2vca})
We first
pin the Dirac metal-insulator transition by determining the opening of the
charge gap $\Delta_{\rm c}$ at $U/t=9.5$ (blue domain in
Fig.~\ref{fig:spec2vca}). Note that this happens at comparably large
$U$, in accordance with the small spectral weight
of the Dirac metal nearby the Fermi level. 
In the infinite coupling limit, the nearest neighbor Heisenberg
term $J=4t^2/U$ dominates the virtual spatial fluctuation processes.
We apply the Weiss field associated with 120$^{\circ}$ N\'eel order and
determine the response of the $\pi$-THM. We find magnetic order
ranging only down to $U/t=13.4$ (red domain in
Fig.~\ref{fig:spec2vca}). This finding is remarkable, as the regular THM, investigated for the same setting, allows for magnetic ordering
to the lower value of $U/t=8.5$. 
(Quantitative deviations from
previous VCA investigations of the regular THM~\cite{PhysRevLett.100.136402} derive from our
consideration of cluster hopping variation.) 
This can be
understood from a strong coupling expansion~\cite{PhysRevB.37.9753}. For the $\pi$-THM, up to
order $t^4/U^3$, we find
\begin{eqnarray} \nonumber
\mathcal{H^{(\text{4})}_{\pi\text{-THM}}} &=& \left( \frac{4t^2}{U} +
  \frac{12 t^4}{U^3}\right) \sum_{\langle ij \rangle} \bs{S}_i\bs{S}_j
+ \frac{12t^4}{U^3}\sum_{\langle\!\langle ij \rangle\!\rangle} \bs{S}_i\bs{S}_j \\ \nonumber
&+& \frac{4 t^4}{U^3} \sum_{\langle\!\langle\!\langle ij
  \rangle\!\rangle\!\rangle} \bs{S}_i\bs{S}_j -\frac{80t^4}{U^3}
\sum_{p}\Big[
\left(\bs{S}_1\bs{S}_2\right)\left(\bs{S}_3\bs{S}_4\right)\Big. \\ 
&+& \Big.\left(\bs{S}_2\bs{S}_3\right)\left(\bs{S}_1\bs{S}_4\right) -
\left(\bs{S}_1\bs{S}_3\right)\left(\bs{S}_2\bs{S}_4\right) \Big]\ , \label{4thorder-spinham}
\end{eqnarray}
where we use the standard notations $\langle ij \rangle$,
$\langle\!\langle ij \rangle\!\rangle$, and $\langle\!\langle\!\langle
ij \rangle\!\rangle\!\rangle$ for first, second, and third nearest
neighbors, and $\sum_p$ indicates the summation over all parallelograms (including different orientations) which consist of two triangles,~
$\begin{picture}(0,0)(0,0)\put(4.1,4.2){\makebox(0,0){$\bigtriangledown$}}
\put(0,2){\makebox(0,0){$\bigtriangleup$}}\end{picture}$~~, where the
long diagonal on the parallelogram is a link between the sites with
indices ``2" and ``4". (Note that spin Hamiltonians including a ring
exchange term such as
\eqref{4thorder-spinham}  were studied
previously on the triangular lattice, albeit for arbitrary coefficients independent of a strong
coupling expansion. There, a limited range
on the Heisenberg exchange coupling was assumed, constraining it up to
second~\cite{bieri-13prl157203} or nearest neigbor~\cite{ken-97zpb485}.)
Comparing $\pi$-THM against the regular THM~\cite{PhysRevB.73.155115,PhysRevLett.105.267204}, one difference is
the reversed sign for the plaquette term coefficient in~\eqref{4thorder-spinham}. We compute the
strength of dimer resonances to investigate the effect of such
higher order contributions. Given a dimer loop, the transition matrix element $|E_{12}|$ between two dimer configurations on that loop determines the energy gain associated with such a resonance. For a loop on a $2\times2$ plaquette, this reads
\begin{equation}
E_{12}=\langle\;\raisebox{-6pt}{\includegraphics[scale=0.8]{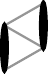}}\;|\mathcal{H}^{\text{(4)}}_{\pi\text{-THM}}|\;\raisebox{-6pt}{\includegraphics[scale=0.8]{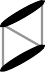}}\;\rangle\;,\label{resonance}
\end{equation}
where
\begin{equation}
|\;\raisebox{-8pt}{\includegraphics[scale=0.8]{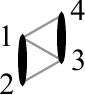}}\;\rangle=\frac{1}{2}(|\hspace*{-2pt}\uparrow_1\downarrow_2\rangle-|\hspace*{-2pt}\downarrow_1\uparrow_2\rangle)(|\hspace*{-2pt}\uparrow_3\downarrow_4\rangle-|\hspace*{-2pt}\downarrow_3\uparrow_4\rangle)\;.\label{singlet}
\end{equation}
In Eq.~\ref{resonance} the Hamiltonian
$\mathcal{H}^{\text{(4)}}_{\pi\text{-THM}}$ is restricted to a
$2\times2$ plaquette. Resonances for longer dimer loops may be
calculated similarly. In Fig.~\ref{fig:dimer}, we show the energy gain
$|E_{12}|$ for resonances on all dimer loops with a length of 4 and 6
lattice spacings. For the shortest 4-site loop (Fig.~\ref{fig:dimer}a), $|E_{12}|$ is larger for the $\pi$-THM than for the regular THM.
This picture diversifies
as we consider longer-loop contributions (Figs.~\ref{fig:dimer}(b)-(d)),
while the general trend from the smallest loop size persists. The
enhanced dimer resonances give a natural explanation for the quick
drop of magnetic order in the $\pi$-THM upon decreasing $U/t$. Whether
a valence bond crystal, i.e. the onset of translational symmetry
breaking, or a spin liquid state might be preferred cannot be inferred from this consideration. (At least note that
the dimer resonance does not significantly drop from a 2-site loop to 6-site loops, which might
suggest a possibly sizable resonance to long-range dimer loops along
the RVB liquid paradigm~\cite{anderson73,rokhsar88,moessner01}; see also Refs.\,\cite{mila-07jpcm145201,vernay-06prb054402}.) Similarly, our analysis does not allow to
determine whether the non-magnetic insulating domain (Fig.~\ref{fig:spec2vca}) is composed out of
one or several distinct paramagnetic phases. What our VCA analysis
does allow to determine, however, is whether collinear magnetic order
can be stabilized, as a recent variational Monte Carlo
study~\cite{bieri-13prl157203} of a similar model as
Eq.\,\eqref{4thorder-spinham} suggests: in the parameter range $5 <
U/t < 15$, we cannot find a stationary point for the collinear magnet,
thus rejecting it as a candidate ground state. Another direction to
further understand the NMI phase within VCA might be to allow
for spatially anisotropic hoppings, where a similarly big NMI domain
has been found previously (see \eg Ref.~\cite{kashima-01jpsp3052}).

{\it Conclusion.} We have proposed the Hubbard
model on the  $\pi$ flux triangular lattice to consitute a paradigmatic scenario for quantum
paramagnets at intermediate coupling. Via VCA, we find a non-magnetic
insulating regime for $9.5 < U/t < 13.4$ framed by a Dirac semi-metal and
120$^{\circ}$ N\'eel order which only establishes itself close to
the strong coupling limit because of significant quantum
fluctuations. The dimer resonances of the $\pi$-THM provide further
support for
its propensity towards quantum paramagnetic phases. 

Several directions
might be interesting to follow up on this work. First, additional methodological
approaches should be applied to further resolve the nature of the
paramagnetic domain in the $\pi$-THM. Second, it is worth
investigating possible experimental realizations in the context of
ultra-cold atomic fermionic gases deposited in optical flux
lattices. In addition, the Klein-transformed Hubbard model derived
from the $\pi$-THM might be applicable to the iridate triangular
compounds where a joint perspective from Heisenberg-Kitaev models and
charge fluctuations due to finite Hubbard $U$ might be
indispensable. Third, from a broader perspective, the Klein duality mapping of
Hubbard models can establish a valuable new tool to derive interesting
connections between different lattice Hamiltonians,
where one model allows to draw implications on the other.

We thank F.\,F.\ Assaad, L.\ Balents, S.\ Bieri, M.\ Voj\-ta, and S.\ Wessel for discussions. We
are grateful to O.~Motrunich for pointing out to us Ref.\,\cite{wen-unpub}. 
SR is supported by DFG-FOR 960, DFG-SPP 1666, DFG-SFB 1143, and by the Helmholtz association through VI-521. JR acknowledges support from the Deutsche Akademie der Naturforscher Leopoldina through grant LPDS 2011-14. ML and RT are
supported by the ERC starting grant TOPOLECTRICS (ERC-StG-Thomale-336012) and
by the National Science Foundation under Grant No.\ NSF PHY11-25915.
We thank the Center for Information Services and High Performance Computing (ZIH) at TU
Dresden for generous allocations of computer time.

\bibliographystyle{prsty}
\bibliography{triflux} 

\begin{thebibliography}{10}

\bibitem{diep04}
{\em Frustrated spin systems}, edited by H.~T. Diep (World Scientific,
  Singapore, 2004).

\bibitem{lacroix}
C. Lacroix, P. Mendels, and F. Mila, {\em Introduction to Frustrated Magnetism:
  Materials, Experiments, Theory} (Springer, Berlin, 2011), Vol.~164.

\bibitem{lees08}
P.~A. Lee, Science {\bf 321},  1306  (2008).

\bibitem{balentsn10}
L. Balents, Nature {\bf 464},  199  (2010).

\bibitem{whites11}
S. Yan, D.~A. Huse, and S.~R. White, Science {\bf 332},  1173  (2011).

\bibitem{jiang-j1j2}
H.-C. Jiang, H. Yao, and L. Balents, Phys. Rev. B {\bf 86},  024424  (2012).

\bibitem{depenbrock-12prl067201}
S. Depenbrock, I.~P. McCulloch, and U. Schollw\"ock, Phys. Rev. Lett. {\bf
  109},  067201  (2012).

\bibitem{suttner-kagome}
R. Suttner, C. Platt, J. Reuther, and R. Thomale, Phys. Rev. B {\bf 89},
  020408  (2014).

\bibitem{PhysRevB.72.045105}
O.~I. Motrunich, Phys. Rev. B {\bf 72},  045105  (2005).

\bibitem{clark-11-j1j2}
B.~K. Clark, D.~A. Abanin, and S.~L. Sondhi, Phys. Rev. Lett. {\bf 107},
  087204  (2011).

\bibitem{meng-10n847}
Z.~Y. Meng, T.~C. Lang, S. Wessel, F.~F. Assaad, and A. Muramatsu, Nature {\bf
  464},  847  (2010).

\bibitem{sorella-12sr992}
S. Sorella, Y. Otsuka, and S. Yunoki, Sci. Rep. {\bf 2},  992  (2012).

\bibitem{PhysRevX.3.031010}
F.~F. Assaad and I.~F. Herbut, Phys. Rev. X {\bf 3},  031010  (2013).

\bibitem{schmidt2014}
D. Ixert, F.~F. Assaad, and K.~P. Schmidt, Phys. Rev. B {\bf 90},  195133
  (2014).

\bibitem{PhysRevLett.110.096402}
S.~R. Hassan and D. S\'en\'echal, Phys. Rev. Lett. {\bf 110},  096402  (2013).

\bibitem{PhysRevB.90.165136}
M. Laubach, J. Reuther, R. Thomale, and S. Rachel, Phys. Rev. B {\bf 90},
  165136  (2014).

\bibitem{abrikosov}
A.~A. Abrikosov and S.~D. Beneslavskii, JETP {\bf 32},  699  (1971).

\bibitem{herbut06prl146401}
I.~F. Herbut, Phys. Rev. Lett. {\bf 97},  146401  (2006).

\bibitem{PhysRevLett.82.3899}
L. Capriotti, A.~E. Trumper, and S. Sorella, Phys. Rev. Lett. {\bf 82},  3899
  (1999).

\bibitem{khaliullin05ptps155}
G. Khaliullin, Prog. Theor. Phys. Suppl. {\bf 160},  155  (2005).

\bibitem{jackeli-09prl017205}
G. Jackeli and G. Khaliullin, Phys. Rev. Lett. {\bf 102},  017205  (2009).

\bibitem{reuther-12prb155127}
J. Reuther, R. Thomale, and S. Rachel, Phys. Rev. B {\bf 86},  155127  (2012).

\bibitem{kimchi-14prb014414}
I. Kimchi and A. Vishwanath, Phys. Rev. B {\bf 89},  014414  (2014).

\bibitem{jaksch-03njp56}
D. Jaksch and P. Zoller, New. J. Phys. {\bf 5},  56  (2003).

\bibitem{dalibard-11rmp1523}
J. Dalibard, F. Gerbier, G. Juzeli\ifmmode~\bar{u}\else \={u}\fi{}nas, and P.
  \"Ohberg, Rev. Mod. Phys. {\bf 83},  1523  (2011).

\bibitem{cooper}
N.~R. Cooper, Phys. Rev. Lett. {\bf 106},  175301  (2011).

\bibitem{struck-12prl225304}
J. Struck, C. \"Olschl\"ager, M. Weinberg, P. Hauke, J. Simonet, A. Eckardt, M.
  Lewenstein, K. Sengstock, and P. Windpassinger, Phys. Rev. Lett. {\bf 108},
  225304  (2012).

\bibitem{wen-unpub}
Y. Zhou and X.-G. Wen, cond-mat/0210662.

\bibitem{PhysRevB.86.140405}
T. Dey, A.~V. Mahajan, P. Khuntia, M. Baenitz, B. Koteswararao, and F.~C. Chou,
  Phys. Rev. B {\bf 86},  140405  (2012).

\bibitem{trebst2017}
M. Becker, M. Herrmanns, B. Bauer, M. Garst, and S. Trebst, arXiv:1409.6972.

\bibitem{li-14arxiv1409.7820}
K. Li, S.-L. Yu, and J.-X. Li, arXiv:1409.7820.

\bibitem{rousochatzakis-12arXiv:1209.5895}
I. Rousochatzakis, U.~K. R{\"o}ssler, J. {van den Brink}, and M. Daghofer,
  arXiv:1209.5895.

\bibitem{potthoff03epjb335}
M. Potthoff, Eur. Phys. J. B {\bf 36},  335  (2003).

\bibitem{potthoff03epjb429}
M. Potthoff, Eur. Phys. J. B {\bf 32},  429  (2003).

\bibitem{PhysRevLett.100.136402}
P. Sahebsara and D. S\'en\'echal, Phys. Rev. Lett. {\bf 100},  136402  (2008).

\bibitem{PhysRevB.37.9753}
A.~H. MacDonald, S.~M. Girvin, and D. Yoshioka, Phys. Rev. B {\bf 37},  9753
  (1988).

\bibitem{bieri-13prl157203}
R. Mishmash, J. Garrison, S. Bieri, and C. Xu, Phys. Rev. Lett. {\bf 111},
  157203  (2013).

\bibitem{ken-97zpb485}
K. Ken and T. Momoi, Z. Phys. B {\bf 103},  485  (1997).

\bibitem{PhysRevB.73.155115}
O.~I. Motrunich, Phys. Rev. B {\bf 73},  155115  (2006).

\bibitem{PhysRevLett.105.267204}
H.-Y. Yang, A.~M. L\"auchli, F. Mila, and K.~P. Schmidt, Phys. Rev. Lett. {\bf
  105},  267204  (2010).

\bibitem{anderson73}
P.~W. Anderson, Mater. Res. Bull. {\bf 8},  153  (1973).

\bibitem{rokhsar88}
D.~S. Rokhsar and S.~A. Kivelson, Phys. Rev. Lett. {\bf 61},  2376  (1988).

\bibitem{moessner01}
R. Moessner and S.~L. Sondhi, Phys. Rev. Lett. {\bf 86},  1881  (2001).

\bibitem{mila-07jpcm145201}
F. Mila, F. Vernay, A. Ralko, F. Becca, P. Fazekas, and K. Penc, J. Phys.:
  Cond. Matt. {\bf 19},  145201  (2007).

\bibitem{vernay-06prb054402}
F. Vernay, A. Ralko, F. Becca, and F. Mila, Phys. Rev. B {\bf 74},  054402
  (2006).

\bibitem{kashima-01jpsp3052}
T. Kashima and M. Imada, J. Phys. Soc. Jpn. {\bf 70},  3052  (2001).

\end{thebibliography}

%
%

\end{document}